\documentstyle[preprint,floats,pra,aps]{revtex}
\tightenlines
\begin{document} \draft

\title{Feynman's Entropy and Decoherence Mechanism}

\author {Y. S. Kim \footnote{electronic address: yskim@physics.umd.edu}}
\address{ Department of Physics, University of Maryland, College Park,
Maryland 20742, U.S.A.}

\maketitle

\begin{abstract}
If we reduce coherence in a given quantum system, the result is
an increase in entropy.  Does this necessarily  convert this quantum
system into a classical system?  The answer to this question is No.
The decrease of coherence means more uncertainty.  This does not
seem to make the system closer to classical system where there are
no uncertainties.  We examine the problem using two coupled harmonic
oscillators where we make observations on one of them while the other
oscillator is assumed to be unobservable or to be in Feynman's rest
of the universe.  Our ignorance about the rest of the universe causes
an increase in entropy.  However, does the system act like a classical
system?  The answer is again No.  When and how does this system
appear like a classical system?  It is shown that this paradox can be
resolved only if measurements are taken along the normal coordinates.
It is also shown that Feynman's parton picture is one concrete physical
example of this decoherence mechanism.

\end{abstract}


\section{Introduction}\label{intro}
According to Feynman, {\it the adventure of our science of physics is a
perpetual attempt to recognize that the different aspects of nature
are really different aspects of the same thing}~\cite{feyaip}.  Feynman
wrote many papers on different subjects of physics, but they are
coming from one paper according to him.  We are not able to combine
all of his papers, but we can consider three of his papers published
during the period 1969-72.

In this report, we would like to consider Feynman's 1969 report
on partons~\cite{fey69}, the 1971 paper he published with his students
on the quark model based on harmonic oscillators~\cite{fkr71}, and
the chapter on density matrix in his
1972 book on statistical mechanics~\cite{fey72}.  In these three
different papers, Feynman deals with three distinct aspects of nature.
We shall see whether Feynman was saying the same thing in these papers.

We approach this problem by developing a mathematical instrument which
can support Feynman's physical ideas spelled out in these seemingly
different papers.  We shall use the mathematics of two coupled
harmonic oscillators~\cite{hkn99ajp}.  The standard procedure for this
two-oscillator system is to separate the Hamiltonian using normal
coordinates.  The transformation to the normal coordinate system becomes
very simple if the two oscillators are identical.  We shall use this
simple mathematics to find a common ground for the above-mentioned
articles written by Feynman.

First, let us look at Feynman's book on statistical mechanics~\cite{fey72},
He makes the following statement about the
density matrix. {\it When we solve a quantum-mechanical problem, what we
really do is divide the universe into two parts - the system in which we
are interested and the rest of the universe.  We then usually act as if
the system in which we are interested comprised the entire universe.
To motivate the use of density matrices, let us see what happens when we
include the part of the universe outside the system}.

In order to see clearly what Feynman had in mind, we use the
above-mentioned couples oscillators.  One of the oscillators is the
world in which we are interested with the other oscillator as the rest
of the universe.
There will be no effects on the first oscillator if the system is
decoupled.  Once coupled, we need a normal coordinate system in order
separate the Hamiltonian.  Then it is straightforward to write down
the wave function of the system.

We shall then observe that the mathematics of this oscillator system is
directly applicable to Lorentz-boosted harmonic oscillator wave
functions, where one variable is the longitudinal coordinate and the
other is the time variable.  The system is uncoupled if the oscillator
wave function is at rest, but the coupling becomes stronger as the
oscillator is boosted to a high-speed Lorentz frame~\cite{knp86}.

We shall then note that for two-body system, such as the hydrogen atom,
there is a time-separation variable which is to be linearly mixed with
the longitudinal space-separation variable.  This space-separation
variable is known as the Bohr radius, but we never talk about the
time-separation variable in the present form of quantum mechanics,
because this time-separation variable belongs to Feynman's rest of the
universe.

If we pretend not to know this time-separation variable, the entropy
of the system will increase when the oscillator is boosted to a
high-speed system~\cite{kiwi90pl}.  Does this increase in entropy correspond to
decoherence?  Not necessarily.  However, in 1969, Feynman observed
the parton effect in which a rapidly moving hadron appears as a
collection of incoherent partons~\cite{fey69}.  This is the decoherence
mechanism we like to discuss in this report.

In Sec.~\ref{coupled}, we review the quantum mechanics of coupled
harmonic oscillators in which one of them corresponds to the world in
which we do physics, and the other in the rest of the universe.  In
Sec.~\ref{restof}, it is shown that the time-separation variable in a
two-body bound state belongs to Feynman's rest of the universe.  It is
shown also that Feynman's oscillator formalism includes this
time-separation variable.
We review in Sec.~\ref{parton} Feynman's parton picture.  Finally,
in Sec. \ref{cohere}, we discuss why partons appear as incoherent
particles.

\section{Coupled Oscillators}\label{coupled}
Two coupled harmonic oscillators serve many different purposes in
physics.  It is well known that this oscillator problem can be
formulated into a problem of a quadratic equation in two variables.
To make a long story short, let us consider a system of two identical
oscillators coupled together by a spring.  The Hamiltonian is
\begin{equation}\label{hamil2}
H = {1\over 2m}\left\{p^{2}_{1} + p^{2}_{2} \right\} +
{1\over 2}\left\{K \left(x_{1}^{2} + x^{2}_{2} \right)
+ 2C x_{1} x_{2} \right\} .
\end{equation}
We are now ready to decouple this Hamiltonian by
making the coordinate rotation:
\begin{equation}\label{normal}
y_{1} = {1 \over \sqrt{2}} \left(x_{1}  - x_{2} \right) , \qquad
y_{2} = {1 \over \sqrt{2}} \left(x_{1}  + x_{2} \right) .
\end{equation}
In terms of this new set of variables, the Hamiltonian can be written as
\begin{equation}\label{eq.6}
H = {1\over 2m} \left\{p^{2}_{1} + p^{2}_{2} \right\} +
{K\over 2}\left\{e^{2\eta} y^{2}_{1} + e^{-2\eta} y^{2}_{2} \right\} ,
\end{equation}
with
\begin{equation}\label{omega}
\exp{(\eta)} = \sqrt{(K + C)/(K - C)} .
\end{equation}
Thus $\eta$ measures the strength of the coupling.
If $y_{1}$ and $y_{2}$ are measured in units of $(mK)^{1/4} $,
the ground-state wave function of this oscillator system is
\begin{equation}\label{wfc}
\psi_{\eta}(x_{1},x_{2}) = \left({1 \over {\pi}}\right)^{1/2}
\exp{\left\{-{1\over 2}(e^{\eta} y^{2}_{1} + e^{-\eta} y^{2}_{2})
\right\} } .
\end{equation}
The wave function is separable in the $y_{1}$ and $y_{2}$ variables.
However, for the variables $x_{1}$ and $x_{2}$, the story is quite
different.

The key question is how quantum mechanical calculations in the world
of the observed variable are affected when we average over the other
variable.  The $x_{2}$ space in this case corresponds to Feynman's
rest of the universe, if we only consider quantum mechanics in the
$x_{1}$ space.  As was discussed in the literature for several
different purposes~\cite{knp86,knp91}, the wave function of
Eq.(\ref{wfc}) can be expanded as
\begin{equation}\label{expan}
\psi_{\eta }(x_{1},x_{2}) = {1 \over \cosh\eta}\sum^{}_{k}
\left(\tanh{\eta \over 2}\right)^{k} \phi_{k}(x_{1}) \phi_{k}(x_{2}) .
\end{equation}
The question then is what lessons we can learn from the situation in
which we average over the $x_{2}$ variable.

In order to study this problem, we use the density matrix.  From this
wave function, we can construct the pure-state density matrix
\begin{equation}
\rho(x_{1},x_{2};x_{1}',x_{2}')
= \psi_{\eta}(x_{1},x_{2})\psi_{\eta}(x_{1}',x_{2}') ,
\end{equation}
If we are not able to make observations on the $x_{2}$, we should
take the trace of the $\rho$ matrix with respect to the $x_{2}$
variable.  Then the resulting density matrix is
\begin{equation}\label{integ}
\rho(x, x') = \int \psi_{\eta}(x,x_{2})
\left\{\psi_{\eta}(x',x_{2})\right\}^{*} dx_{2} .
\end{equation}
We have simplicity replaced $x_{1}$ and $x'_{1}$ by $x$ and $x'$
respectively.
If we perform the integral of Eq.(\ref{integ}), the result is
\begin{equation}\label{dmat}
\rho(x,x') = \left({1 \over \cosh(\eta/2)}\right)^{2}
\sum^{}_{k} \left(\tanh{\eta \over 2}\right)^{2k}
\phi_{k}(x)\phi^{*}_{k}(x') ,
\end{equation}
which leads to $Tr(\rho) = 1$.  It is also straightforward to compute
the integral for $Tr(\rho^{2})$.  The calculation leads to
\begin{equation}
Tr\left(\rho^{2} \right)
= \left({1 \over \cosh(\eta/2)}\right)^{4}
\sum^{}_{k} \left(\tanh{\eta \over 2}\right)^{4k} .
\end{equation}
The sum of this series is $(1/\cosh\eta)$, which is smaller than one
if the parameter $\eta$ does not vanish.

This is of course due to the fact that we are averaging over the $x_{2}$
variable which we do not measure.  The standard way to measure this
ignorance is to calculate the entropy defined as
\begin{equation}
S = - Tr\left(\rho \ln(\rho) \right) ,
\end{equation}
where $S$ is measured in units of Boltzmann's constant.  If we use the
density matrix given in Eq.(\ref{dmat}), the entropy becomes
\begin{equation}
S = 2 \left\{\cosh^{2}\left({\eta \over 2}\right)
\ln\left(\cosh{\eta \over 2}\right) -
\sinh^{2}\left({\eta \over 2}\right)
\ln\left(\sinh{\eta \over 2} \right)\right\} .
\end{equation}
This expression can be translated into a more familiar form if
we use the notation
\begin{equation}
\tanh{\eta \over 2} = \exp\left(-{\hbar\omega \over kT}\right) ,
\end{equation}
where $\omega$ is given in Eq.(\ref{omega})~\cite{hkn89pl}.

It is known in the literature that this rise in entropy and temperature
causes the Wigner function to spread wide in phase space causing an
increase of uncertainty~\cite{hkn99ajp}.  Certainly, we cannot reach a
classical limit by increasing the uncertainty.  On the other hand, we
are accustomed to think this entropy increase has something to do with
decoherence, and we are also accustomed to think the lack of coherence
has something to do with a classical limit.  Are they compatible?  We
thus need a new vision in order to define precisely the word
``decoherence.''

\section{Time-separation Variable in Feynman's Rest of
the Universe}\label{restof}
Quantum field theory has been quite successful in terms of
perturbation techniques in quantum electrodynamics.  However, this
formalism is basically based
on the S matrix for scattering problems and useful only for physically
processes where free a set of particles becomes another set of free
particles after interaction.  Quantum field theory does not address
the question of localized probability distributions and their
covariance under Lorentz transformations.
The Schr\"odinger quantum mechanics of the hydrogen atom deals with
localized probability distribution.  Indeed, the localization condition
leads to the discrete energy spectrum.  Here, the uncertainty relation
is stated in terms of the spatial separation between the proton and
the electron.  If we believe in Lorentz covariance, there must also
be a time separation between the two constituent particles.

Before 1964~\cite{gell64}, the hydrogen atom was used for
illustrating bound states.  These days, we use hadrons which are
bound states of quarks.  Let us use the simplest hadron consisting of
two quarks bound together an attractive force, and consider their
space-time positions $x_{a}$ and $x_{b}$, and use the variables
\begin{equation}
X = (x_{a} + x_{b})/2 , \qquad x = (x_{a} - x_{b})/2\sqrt{2} .
\end{equation}
The four-vector $X$ specifies where the hadron is located in space and
time, while the variable $x$ measures the space-time separation
between the quarks.  According to Einstein, this space-time separation
contains a time-like component which actively participates as can be
seen from
\begin{equation}\label{boostm}
\pmatrix{z' \cr t'} = \pmatrix{\cosh \eta & \sinh \eta \cr
\sinh \eta & \cosh \eta } \pmatrix{z \cr t} ,
\end{equation}
when the hadron is boosted along the $z$ direction.
In terms of the light-cone variables defined as~\cite{dir49}
\begin{equation}
u = (z + t)/\sqrt{2} , \qquad v = (z - t)/\sqrt{2} .
\end{equation}
The boost transformation of Eq.(\ref{boostm}) takes the form
\begin{equation}\label{lorensq}
u' = e^{\eta } u , \qquad v' = e^{-\eta } v .
\end{equation}
The $u$ variable becomes expanded while the $v$ variable becomes
contracted.

Does this time-separation variable exist when the hadron is at rest?
Yes, according to Einstein.  In the present form of quantum mechanics,
we pretend not to know anything about this variable.  Indeed, this
variable belongs to Feynman's rest of the universe.  In this report,
we shall see the role of this time-separation variable in decoherence
mechanism.

Also in the present form of quantum mechanics, there is an uncertainty
relation between the time and energy variables.  However, there are
no known time-like excitations.  Unlike Heisenberg's
uncertainty relation applicable to position and momentum, the time and
energy separation variables are c-numbers, and we are not allowed to
write down the commutation relation between them.  Indeed, the
time-energy uncertainty relation is a c-number uncertainty
relation~\cite{dir27}.

How does this space-time asymmetry fit into the world of
covariance~\cite{kn73}.  This question was
studied in depth by the present author and his collaborators.  The
answer is that Wigner's $O(3)$-like little group is not a
Lorentz-invariant symmetry, but is a covariant symmetry~\cite{wig39}.
It has been shown that the time-energy uncertainty applicable to the
time-separation variable fits perfectly into the $O(3)$-like symmetry
of massive relativistic particles~\cite{knp86}.

The c-number time-energy uncertainty relation allows us to write down
a time distribution function without excitations~\cite{knp86}.
If we use Gaussian forms for both space and time distributions, we
can start with the expression
\begin{equation}\label{ground}
\left({1 \over \pi} \right)^{1/2}
\exp{\left\{-{1 \over 2}\left(z^{2} + t^{2}\right)\right\}}
\end{equation}
for the ground-state wave function.  What do Feynman {\it et al.}
say about this oscillator wave function?

In his classic 1971 paper~\cite{fkr71}, Feynman {\it et al.} start
with the following Lorentz-invariant differential equation.
\begin{equation}\label{osceq}
{1\over 2} \left\{x^{2}_{\mu} -
{\partial^{2} \over \partial x_{\mu }^{2}}
\right\} \psi(x) = \lambda \psi(x) .
\end{equation}
This partial differential equation has many different solutions
depending on the choice of separable variables and boundary conditions.
Feynman {\it et al.} insist on Lorentz-invariant solutions which are
not normalizable.  On the other hand, if we insist on normalization,
the ground-state wave function takes the form of Eq.(\ref{ground}).
It is then possible to construct a representation of the
Poincar\'e group from the solutions of the above differential
equation~\cite{knp86}.  If the system is boosted, the wave function
becomes
\begin{equation}\label{eta}
\psi_{\eta }(z,t) = \left({1 \over \pi }\right)^{1/2}
\exp\left\{-{1\over 2}\left(e^{-2\eta }u^{2} +
e^{2\eta}v^{2}\right)\right\} .
\end{equation}
This wave function becomes Eq.(\ref{ground}) if $\eta$ becomes zero.
The transition from Eq.(\ref{ground}) to Eq.(\ref{eta}) is a
squeeze transformation.  The wave function of Eq.(\ref{ground}) is
distributed within a circular region in the $u v$ plane, and thus in
the $z t$ plane.
On the other hand, the wave function of Eq.(\ref{eta}) is distributed
in an elliptic region with the light-cone axes as the major and
minor axes respectively.  If $\eta$ becomes very large, the wave
function becomes concentrated along one of the light-cone axes.
Indeed, the form given in Eq.(\ref{eta}) is a Lorentz-squeezed wave
function.

It is interesting to note that the Lorentz-invariant differential
equation of Eq.(\ref{osceq}) contains the time-separation variable
which belongs to Feynman's rest of the universe.  Furthermore, the
wave function of Eq.(\ref{ground}) is identical to that of
Eq.(\ref{wfc}) for the coupled oscillator system, if the variables
$z$ and $t$ are replaced $x_{1}$ and $x_{2}$ respectively.
Thus the entropy increase due to the unobservable $x_{2}$ variable is
applicable to the unobserved time-separation variable $t$.

\section{Feynman's Parton Picture}\label{parton}

It is a widely accepted view that hadrons are quantum bound states
of quarks having localized probability distribution.  As in all
bound-state cases, this localization condition is responsible for
the existence of discrete mass spectra.  The most convincing evidence
for this bound-state picture is the hadronic mass spectra which are
observed in high-energy laboratories~\cite{fkr71,knp86}.
However, this picture of bound states is applicable only to observers
in the Lorentz frame in which the hadron is at rest.  How would the
hadrons appear to observers in other Lorentz frames?  To answer this
question, can we use the picture of Lorentz-squeezed hadrons discussed
in Sec.~\ref{restof}.

The radius of the proton is $10^{-5}$ of that of the hydrogen atom.
Therefore, it is not unnatural to assume that the proton has a point
charge in atomic physics.  However, while carrying out experiments on
electron scattering from proton targets, Hofstadter in 1955 observed
that the proton charge is spread out~\cite{hofsta55}.
In this experiment, an electron emits a virtual photon, which
then interacts with the proton.  If the proton consists of quarks
distributed within a finite space-time region, the virtual photon will
interact with quarks which carry fractional charges.  The scattering
amplitude will depend on the way in which quarks are distributed
within the proton.  The portion of the scattering amplitude which
describes the interaction between the virtual photon and the proton
is called the form factor.

Although there have been many attempts to explain this phenomenon
within the framework of quantum field theory, it is quite natural
to expect that the wave function in the quark model will describe
the charge distribution.  In high-energy experiments, we are dealing
with the situation in which the momentum transfer in the scattering
process is large.  Indeed, the Lorentz-squeezed wave functions lead
to the correct behavior of the hadronic form factor for large
values of the momentum transfer~\cite{fuji70}.

Furthermore, in 1969, Feynman observed that a fast-moving hadron
can be regarded as a collection of many ``partons'' whose properties
do not appear to be quite different from those of the
quarks~\cite{fey69}.  For example, the number of quarks inside a
static proton is three, while the number of partons in a rapidly
moving proton appears to be infinite.  The question then is how
the proton looking like a bound state of quarks to one observer
can appear different to an observer in a different Lorentz frame?
Feynman made the following systematic observations.

\begin{itemize}

\item[a.]  The picture is valid only for hadrons moving with
  velocity close to that of light.

\item[b.]  The interaction time between the quarks becomes dilated,
   and partons behave as free independent particles.

\item[c.]  The momentum distribution of partons becomes widespread as
   the hadron moves fast.

\item[d.]  The number of partons seems to be infinite or much larger
    than that of quarks.

\end{itemize}

\noindent Because the hadron is believed to be a bound state of two
or three quarks, each of the above phenomena appears as a paradox,
particularly b) and c) together.

In order to resolve this paradox, let us write down the
momentum-energy wave function corresponding to Eq.(\ref{eta}).
If the quarks have the four-momenta $p_{a}$ and $p_{b}$, we can
construct two independent four-momentum variables~\cite{fkr71}
\begin{equation}
P = p_{a} + p_{b} , \qquad q = \sqrt{2}(p_{a} - p_{b}) .
\end{equation}
The four-momentum $P$ is the total four-momentum and is thus the
hadronic four-momentum.  $q$ measures the four-momentum separation
between the quarks.  Their light-cone variables are
\begin{equation}\label{conju}
q_{u} = (q_{0} - q_{z})/\sqrt{2} ,  \qquad
q_{v} = (q_{0} + q_{z})/\sqrt{2} .
\end{equation}
The resulting momentum-energy wave function is
\begin{equation}\label{phi}
\phi_{\eta }(q_{z},q_{0}) = \left({1 \over \pi }\right)^{1/2}
\exp\left\{-{1\over 2}\left(e^{-2\eta}q_{u}^{2} +
e^{2\eta}q_{v}^{2}\right)\right\} .
\end{equation}
Because we are using here the harmonic oscillator, the mathematical
form of the above momentum-energy wave function is identical to that
of the space-time wave function.  The Lorentz squeeze properties of
these wave functions are also the same.  This aspect of the squeeze
has been exhaustively discussed in the
literature~\cite{knp86,kn77par,kim89}.

When the hadron is at rest with $\eta = 0$, both wave functions
behave like those for the static bound state of quarks.  As $\eta$
increases, the wave functions become continuously squeezed until
they become concentrated along their respective positive
light-cone axes.  Let us look at the z-axis projection of the
space-time wave function.  Indeed, the width of the quark distribution
increases as the hadronic speed approaches that of the speed of
light.  The position of each quark appears widespread to the observer
in the laboratory frame, and the quarks appear like free particles.

The momentum-energy wave function is just like the space-time wave
function.  The longitudinal momentum distribution becomes wide-spread
as the hadronic speed approaches the velocity of light.  This is in
contradiction with our expectation from nonrelativistic quantum
mechanics that the width of the momentum distribution is inversely
proportional to that of the position wave function.  Our expectation
is that if the quarks are free, they must have their sharply defined
momenta, not a wide-spread distribution.

However, according to our Lorentz-squeezed space-time and
momentum-energy wave functions, the space-time width and the
momentum-energy width increase in the same direction as the hadron
is boosted.  This is of course an effect of Lorentz covariance.
This indeed is the key to the resolution of the quark-parton
paradox~\cite{knp86,kn77par}.

\section{Decoherence in the Parton Picture}\label{cohere}
The most puzzling problem in the parton picture is that partons in
the hadron appear as incoherent particles, while quarks are coherent
when the hadron is at rest.  Does this mean that the coherence is
destroyed by the Lorentz boost?   The answer is NO, and here is the
resolution to this puzzle.

When the hadron is boosted, the hadronic matter becomes squeezed and
becomes concentrated in the elliptic region along the positive
light-cone axis.  The length of the major axis becomes expanded by
$e^{\eta}$, and the minor axis is contracted by $e^{\eta}$.

This means that the interaction time of the quarks among themselves
become dilated.  Because the wave function becomes wide-spread, the
distance between one end of the harmonic oscillator well and the
other end increases.  This effect, first noted by Feynman~\cite{fey69},
is universally observed in high-energy hadronic experiments.  The
period is oscillation is increases like $e^{\eta}$.

On the other hand, the interaction time with
the external signal, since it is moving in the direction opposite to
the direction of the hadron, it travels along the negative light-cone
axis.  If the hadron contracts along the negative light-cone axis, the
interaction time decreases by $e^{-\eta}$.  The ratio of the interaction
time to the oscillator period becomes $e^{-2\eta}$.  The energy of each
proton coming out of the Fermilab accelerator is $900 GeV$.  This leads
the ratio to $10^{-6}$.  This is indeed a small number.  The external
signal is not able to sense the interaction of the quarks among
themselves inside the hadron.

Indeed, Feynman's parton picture is one concrete physical example
where the decoherence effect is observed.  As for the entropy, the
time-separation variable belongs to the rest of the universe.  Because
we are not able to observe this variable, the entropy increases
as the hadron is boosted to exhibit the parton effect.  The
decoherence is thus accompanied by an entropy increase.

Let us go back to the coupled-oscillator system.  The light-cone
variables in Eq.(\ref{eta}) correspond to the normal coordinates in
the coupled-oscillator system given in Eq.(\ref{normal}).  According
to Feynman's parton picture, the decoherence mechanism is determined
by the ratio of widths of the wave function along the two normal
coordinates.

\end{document}